\documentclass[superscriptaddress,nofootinbib,twocolumn]{revtex4-1}

\usepackage[utf8]{inputenc}

\usepackage[colorlinks]{hyperref}
\hypersetup{
     colorlinks   = true,
     citecolor    = red,
	linkcolor=blue
}

\usepackage{bbm}
\usepackage{amsfonts}
\usepackage{mathrsfs}

\usepackage{amsmath,amssymb}

\usepackage{epsfig}
\usepackage{graphicx}               
\usepackage{url}
\usepackage{hyperref}
\usepackage{float}
\usepackage{soul}
\usepackage{pstricks}
\usepackage{color}
\usepackage{multirow}

\makeatletter
\def\simgt{\mathrel{\lower2.5pt\vbox{\lineskip=0pt\baselineskip=0pt
           \hbox{$>$}\hbox{$\sim$}}}}
\def\simlt{\mathrel{\lower2.5pt\vbox{\lineskip=0pt\baselineskip=0pt
           \hbox{$<$}\hbox{$\sim$}}}}
\makeatother

\def\mysection#1{{\bf #1.} }

\newcommand{\be}{\begin{equation}}
\newcommand{\ee}{\end{equation}}
\newcommand{\bea}{\begin{eqnarray}}
\newcommand{\eea}{\end{eqnarray}}
\newcommand{\beq}{\begin{eqnarray}}
\newcommand{\eeq}{\end{eqnarray}}

\DeclareMathOperator{\sinc}{sinc}

\def\lsim{\mathrel{\rlap{\lower4pt\hbox{\hskip1pt$\sim$}}
     \raise1pt\hbox{$<$}}}         
\def\gsim{\mathrel{\rlap{\lower4pt\hbox{\hskip1pt$\sim$}}
     \raise1pt\hbox{$>$}}}         

\begin{document}

\title{On Vacuum Fluctuations in Quantum Gravity and Interferometer Arm Fluctuations}
\author{Kathryn M. Zurek}
\affiliation{Walter Burke Institute for Theoretical Physics \\ California Institute of Technology, Pasadena, CA USA }

\begin{abstract}

We propose a simple model of spacetime vacuum fluctuations motivated by AdS/CFT, where the vacuum is described by a thermal density matrix, $\rho = \frac{e^{-K}}{\mbox{Tr}(e^{-K})}$ with $K$ the modular Hamiltonian.  In AdS/CFT, both the expectation value of $K$ and its fluctuations $\langle \Delta K^2 \rangle$ have been calculated; both obey an area law identical to the Bekenstein-Hawking area law of black hole mechanics: $\langle K \rangle = \langle \Delta K^2 \rangle = \frac{A}{4 G_N}$, where $A$ is the area of an (extremal) entangling surface.  It has also been shown that $\Delta K$ gravitates in AdS, and hence generates metric fluctuations.  These theoretical results are intriguing, but it is not known how to precisely extend such ideas about holographic quantum gravity to ordinary flat space. We take the approach of considering whether experimental signatures in metric fluctuations could determine properties of the vacuum of quantum gravity in flat space.  In particular, we propose a theoretical model motived by the AdS/CFT calculations that reproduces the most important features of modular Hamiltonian fluctuations; the model consists of a high occupation number bosonic degree of freedom.  We show that if this theory couples through ordinary gravitational couplings to the mirrors in an interferometer with strain sensitivity similar to what will be available for gravitational waves, vacuum fluctuations could be observable.

\end{abstract}

\maketitle

\mysection{Overview}
\label{sec:intro}  
One of the most important open problems in fundamental physics is bridging the theoretical and observational chasm between quantum mechanics and gravity, to understand quantum gravity.  The divide can be understood at a basic level by considering the length and time scales of gravity, which are derived from the Newton constant, $l_p = \sqrt{8 \pi G \hbar / c^3} \simeq 10^{-34} \mbox{ m} ,~t_p = l_p/c \simeq 10^{-43} \mbox{ s}$.\footnote{For the remainer of this letter we will take $\hbar = c = 1$.}  From the point of view of quantum mechanics and Effective Field Theory (EFT), these numbers represent the time and length scales of quantum fluctuations.  They are far out of reach of observational capabilities, and neatly encapsulate why the quantum effects of gravity have never been probed.

On the other hand, theoretical progress, mostly in the context of AdS/CFT and the black hole information paradox, suggests that non-locality and entanglement plays an important role in the quantum theory of gravity (see Ref.~\cite{VanRaamsdonk:2016exw} for a pedagogical review and references therein).   The simplest way to see why quantum gravity may be highly non-local, and hence why the na\"ive expectations of EFT may break down, is through holography.  Quantum Field Theory (QFT) dictates that the number of degrees of freedom in a spacetime volume scales with the volume (in units of $l_p$).  Holography, on the other hand, says that the degrees-of-freedom of a spacetime volume scales with the area of the surface bounding that volume.  As a result,  this implies that QFT grossly over-counts the number of degrees-of-freedom when gravity is involved, suggesting that in any EFT description of spacetime, there should be long range correlations between the degrees-of-freedom.  It is such long range correlations that we seek to model in this letter, where we call each spacetime degree-of-freedom a {\em pixel}.

Our ability to quantify spacetime fluctuations from pixels relies on theoretical progress made recently in connecting the holographic notion of spacetime to entanglement.  Not surprisingly, the connection between geometry, quantum information (QI) and holography has been formulated most clearly and rigorously in the context of AdS/CFT \cite{Ryu:2006ef,Ryu:2006bv}, though there are reasons to think that the connection appears in generic spacetimes \cite{Srednicki:1993im,Callan:1994py,Jacobson:1995ab,Jacobson:2015hqa}.   
In the context of AdS/CFT, the {\em geometry-QI-holography} connection is encapsulated most prominently in the relation
\beq
\beta \langle K \rangle = S_{\rm ent} = \frac{A(\Sigma)}{4 G}.
\label{eq:K}
\eeq
Let us discuss each part of this equality. $K$ is a measure of the total energy in a spacetime volume called the modular Hamiltonian of the boundary CFT ($K \equiv \int T_{\mu \nu} \xi^\mu dV^\nu$ with $T_{\mu \nu}$ the CFT stress tensor, $\xi^\mu$ the conformal Killing vector on the boundary and $dV^\nu$ an infinitesimal volume element); since energy directly sources the metric through the Einstein equation, it represents {\em geometry}.
The entanglement entropy $S_{\rm ent}$ quantifies mixing of the vacuum state $\rho_\beta$ across an entangling boundary and represents the {\em QI} part of the correspondence: $S_{\rm ent}= -\mbox{Tr}\rho_\beta\log\rho_\beta$, where the entangling surface $\Sigma$ has area $A(\Sigma)$ and an associated inverse temperature $\beta = 2\pi R$, where $R$ is the AdS curvature.\footnote{We have reintroduced the temperature; it can be reabsorbed in the definition of $K$ as in the abstract, but we will find it convenient to leave it explicit.} It was remarkably shown \cite{Ryu:2006bv} that there is a special surface in AdS, called the RT surface, where the entanglement entropy of the vacuum state has precisely the form of the Bekenstein-Hawking area law for the entropy of black hole horizons; this represents the {\em holographic} part of the connection. 

Further, by calculating fluctuations of the modular Hamiltonian (and of the entanglement entropy), one can compute fluctuations in the bulk (AdS) geometry.  The fluctuations of the modular Hamiltonian have been calculated in the AdS bulk \cite{Verlinde:2019ade}, with the result
\beq
\beta^2 \langle \Delta K^2 \rangle = \frac{A(\Sigma)}{4 G}.
\label{eq:DelK}
\eeq
With suitable identifications, this result agrees with an equivalent boundary calculation \cite{Perlmutter:2013gua}, and with the ``capacity of entanglement'' \cite{Nakaguchi:2016zqi,deBoer:2018mzv}.  This implies that, when restricted to a finite part of the spacetime (defined by a causal diamond with bifurcate horizon $\Sigma$), the vacuum has energy fluctuations $\Delta K \neq 0$.

Eqs.~(\ref{eq:K}),~(\ref{eq:DelK}) together encourage an interpretation of spacetime as bits of information, with the number ${\cal N}$ of degrees-of-freedom in a given volume bounded by a surface of area $A$ given by the entanglement entropy
\beq
S_{\rm ent} = {\cal N} = \frac{A}{4 G}.
\label{eq:N}
\eeq
This result can be interpreted in light of the fact that  the vacuum state of any QFT, restricted to a causal diamond, is given by a thermal density matrix \cite{Casini:2011kv}
\beq
\rho_\beta = \frac{e^{-\beta K}}{Z_\beta}.
\label{eq:densmat}
\eeq
For example, in a high-temperature system with Maxwell-Boltzmann statistics, the root-mean-square fluctuations of these ${\cal N}$ degrees-of-freedom is given by
\beq
\frac{\sqrt{\langle \Delta K^2\rangle}}{\langle K \rangle} = \frac{1}{\sqrt{\cal N}},
\label{eq:sqrtN}
\eeq 
in agreement with Eqs.~(\ref{eq:K}),~(\ref{eq:DelK}).
It was shown in Ref.~\cite{Verlinde:2019ade} that the fluctuations in the modular Hamiltonian, $\Delta K$, gravitate and hence, for certain observers, behave like a mass, sourcing metric fluctuations.  

It is not precisely known whether the vacuum state of quantum gravity in ordinary flat space, when restricted to a causal diamond, can be described by Eq.~(\ref{eq:densmat}), and whether the results derived for AdS/CFT apply to the Universe we observe.  There are reasons to think that entropy, entanglement and their connection to geometry are very generic concepts that apply to any spacetime, including ours.  Such ideas underly currents dating more than twenty years on the entropic and holographic nature of spacetime \cite{Jacobson:1995ab,Verlinde:2010hp,Banks:2011av}.  For example, it was shown in Ref.~\cite{Jacobson:2015hqa} that taking fixed volume variations of the first law of entanglement gives rise to the Einstein Equations.

In this letter we approach quantum gravity in flat space in terms of observational signatures derived from a model motivated by the known AdS/CFT results.  If vacuum fluctuations could be observed in an experiment consistent with an entropic or thermal nature of the vacuum state, this would be a leap forward in our understanding of quantum gravity.  To this end, we propose a simple model for the degrees-of-freedom in the density matrix, which we call {\em pixellons} because the excitations are associated with each holographic pixel of a volume bounded by an entangling surface of area $A$.  More specifically, we consider whether fluctuations in the spacetime degrees-of-freedom of the density matrix Eq.~(\ref{eq:densmat}) could be observable as fluctuations in the arm length of an interferometer.  The two arms of an interferometer mark out a (spherically symmetric) volume of spacetime with the beamsplitter at the center of the volume and the mirrors on the surface with area $A$; the interferometer measures the geometric fluctuations in this volume.  

The motivation for our pixellon {\em Ansatz} is as follows.  Based on the discussion of Eqs.~(\ref{eq:K})-(\ref{eq:N}), we interpret the spacetime volume bounded by a surface of area $A$ as having ${\cal N}$ bits with total energy $\langle K \rangle$.  The energy per bit is then $\omega \sim \beta^{-1} \sim 1/L$.  However, it was shown in Ref.~\cite{Verlinde:2019ade} that $\langle K \rangle$ itself does not gravitate; rather the {\em fluctuations} $\Delta K$ gravitate.  From this point of view, $\langle K \rangle$ should be treated as a chemical potential counting the background degrees-of-freedom, and the energy per excitation is $\beta \omega = \Delta K/K = 1/\sqrt{\cal N} \ll 1$.  We assume the low-energy excitations will be bosonic; this is appropriate because, as we will see below, they are associated with a gravitational potential.  Because the energy of these bosonic degrees-of-freedom is so low, $\beta \omega \ll 1$, they will form a high-occupation-number bosonic state.  We further show that such low-energy bosonic excitations of the vacuum, when gravitationally coupled to test masses, may give rise to observably large fluctuations of the mirror positions in an interferometer.

The detailed outline of our proposal is as follows.  In the next section we introduce the bosonic excitations as vacuum fluctuations, and we suggest that these bosonic fluctuations be associated with a scalar gravitational potential.  In the following section, we propose that these scalar degrees-of-freedom have a Bose-Einstein density-of-states and a high occupation number due to their low energy.  Then, we gravitationally couple the pixellons to a test mass whose position fluctuates due to the boson fluctuations.  We utilize the Feynman-Vernon influence functional--a path integral realization of the fluctuation-dissipation theorem--to compute the size of those fluctuations.  We will conclude that vacuum fluctuations from a thermal density matrix can give rise to interferometer mirror position fluctuations that appear as noise with a peculiar angular correlation, a smoking gun signature.  Note that while the experimental system of interest for measuring metric fluctuations, an interferometer, and the motivation derived from the discussion surrounding Eqs.~(\ref{eq:K})-(\ref{eq:sqrtN}), is the same as Ref.~\cite{Verlinde:2019ade,Verlinde:2019xfb}, the models operationally share no overlap.  We will nevertheless find a very similar effect, perhaps suggesting dual languages to describe the vacuum state of quantum gravity. 

\mysection{Pixellons and Vacuum Fluctuations}  
\label{sec:GravCond} 
As outlined in the introduction, excitations of the degrees-of-freedom, associated with the entanglement entropy of a finite volume of space, we refer to as {\em pixellons}.  We expect that pixellons are complicated non-linear states of all the degrees-of-freedom available in the complete theory.  Our goal is to describe a consistent low-energy theory based on the quantum information theoretic picture presented in the introduction.

Our starting point is energy fluctuations given by Eq.~\eqref{eq:sqrtN}, which we associate with pixellons.
The energy fluctuations in a volume of radius $L$ source a potential
\beq
2 \phi = \sqrt{\frac{\alpha}{\cal N}} = \frac{ \bar l_p}{ 4\pi   L},
\label{eq:pot}
\eeq
where $l_p^2 \equiv 8 \pi G_N$ and $\bar l_p^2 \equiv \frac{\alpha}{2} l_p^2 $.  Here we have allowed for an ${\cal O}(1)$ number $\alpha$ in the relation $\beta^2\langle \Delta K^2 \rangle = \alpha \beta \langle K \rangle$ ($\alpha = 1$ in AdS/CFT) as a generalization of Eq.~\eqref{eq:sqrtN}. This gravitational potential should be interpreted as the root-mean-square (RMS) fluctuations in the potential about the vacuum state, for which $\langle \phi \rangle = 0$.  Note that this potential is in agreement with the results of Ref.~\cite{Verlinde:2019xfb}, where $\alpha = 2$ (in four dimensions) was derived by computing mass fluctuations thermodynamically and then extracting the gravitational potential from a topological black hole foliation of 4-d Minkowski space.  Next we consider how to represent this gravitational potential as a bosonic degree-of-freedom.

\mysection{Thermal Representation of Vacuum}
\label{sec:ThermRep}
The effects of interest occur because of the finite volume, and hence finite ${\cal N}$, under observation; in the limit ${\cal N} \rightarrow \infty$, fluctuations vanish. 
We divide our space into two regions, $A$ (which is interrogated by an experiment), and $\tilde A$ (outside of the experimental apparatus).  We assume that the vacuum state of the quantized Einstein-Hilbert action, restricted to the subspace $A$, has a density matrix given by Eq.~(\ref{eq:densmat}), such that $\rho_A = \rho_\beta$.  This assumption is supported by Jacobson's derivation \cite{Jacobson:2015hqa} of the Einstein Equation from the entanglement entropy of a diamond utilizing the density matrix in Eq.~(\ref{eq:densmat}).

Furthermore, we are seeking to describe the excitations associated with the fluctuating scalar potential in Eq.~(\ref{eq:pot}).  Accordingly, we assume that there is an algebra of $N = \sqrt{\cal N}$ bosonic degrees-of-freedom to describe the pixellon excitations,
with thermal density matrix given by
\beq
\mbox{Tr}(\rho_\beta a_i^\dagger a_i) = \frac{1}{e^{\beta (\epsilon_i-\mu)} - 1},
\eeq
where the bosonic creation and annihilation operators satisfy the usual commutation relations
\beq
[a_i,a_{i'}^\dagger] = \delta_{i,i'},
\eeq
with $i,~i'$ running from 1 to $N$.   The Hilbert space of these operators has been cut-off by the finite size of the causal diamond. 
If $\epsilon_i-\mu$ represents the energy per degree-of-freedom of excitations of the vacuum state, we have 
\beq
\beta (\epsilon_i - \mu) = \beta \frac{K - \langle K \rangle}{{\cal N}}.
\eeq
 Since $\beta^2 \langle \Delta K^2 \rangle = \alpha \cal N$, and $\beta \sim L$, we immediately learn that the amplitude of a mass fluctuation is Planckian, $\sqrt{\langle \Delta K^2 \rangle} \sim 1/l_p$.  This is a large mass fluctuation, but it is divided over ${\cal N}$ pixels such that the typical energy per degree-of-freedom, relative to the momentum $p \sim 1/\beta$, is
\beq
\beta (\epsilon_{\rm pix}-\mu) \sim \frac{1}{\sqrt{\cal N}} \ll 1.
\label{eq:epspix}
\eeq
To fully describe the distribution, we need the momentum dependence of $\beta (\epsilon_{\rm pix}-\mu)$.  
To obtain this, we employ an {\em Ansatz} motivated by the observation that we produce the scaling in Eq.~(\ref{eq:epspix}) if we postulate   
\beq
\epsilon_{\rm pix} - \mu \equiv  \frac{p^2}{2 m_{\rm pix}} ,
\eeq 
with $m_{\rm pix} \sim 1/l_p$.    
We thus have a pixellon density of states given by
\beq
\rho_{\rm pix}(p) = \frac{1}{e^{\beta \frac{p^2}{2 m_{\rm pix}}}-1}.
\label{eq:rhopix}
\eeq
Because of the extremely low energy of the bosonic degrees-of-freedom, the occupation number of the excitations is high:
\beq
\rho_{\rm pix} \approx \frac{2 m_{\rm pix}}{\beta p^2} \sim \sqrt{\cal N}.
\label{eq:rhopixapprox}
\eeq
This high occupation number will amplify the effective gravitational coupling of a massive object to the pixellons.   Note that while Eq.~(\ref{eq:rhopix}) is the density of states for the collective pixellon modes, one should not identify the dispersion of an individual pixellon as $\omega_p = p^2/2 m_{\rm pix}$; the individual pixellon dispersion will be read off from the action Eq.~(\ref{eq:linearAction}) below.  
We also note in passing that, while the context is different, our hypothesis shares features in common with Ref.~\cite{Dvali:2011aa} for black holes, a connection that might be interesting to explore in the future.

\mysection{Length from Vacuum Fluctuations}
For our purposes, the most important question is whether such excitations are observable. The Ansatz of a pixellon excitations, $\phi$, with density of states given by Eq.~(\ref{eq:rhopix}), coupled gravitationally to the mirror system of an interferometer, is ideally suited to the Feynman-Vernon (FV) influence functional \cite{Feynman:1963fq}.   Refs.~\cite{Parikh:2020nrd,Parikh:2020fhy} have utilized the FV set-up to compute the impact of graviton fluctuations coupled to an interferometer mirror; they find that the mirror position fluctuations are Planckian ({\em i.e.} unobservably small) for the vacuum state.  In this section we utilize similar methods, but focus on the pixellon vacuum fluctuations, and find the effects are large, amplified by the high occupation number of the pixellon.

We will consider a single arm of a simple Michelson interferometer as the test system, with the far mirror of mass $m_M$ at coordinate ${\bf M}$, with equilibrium position a distance $L$ from a beam splitter at position ${\bf B}$.    Our starting point is the mode decomposition
$
h_{ij}(t,{\bf x}) =  \int \frac{d^3 k}{(2\pi)^3} \phi(t,{\bf k}) e^{i {\bf k} \cdot {\bf x}} \epsilon_{ij}^s,
$ 
where $\epsilon_{ij}^s$ is the graviton polarization vector.  We choose TT gauge where $\epsilon_{ij}^s \epsilon^{ij}_t = 2 \delta^{s}_t$, and the metric fluctuations are transverse to the interferometer arm.  $\phi(t,{\bf k})$ will be identified with the pixellon momentum modes.
The action for the mirror and metric is
\begin{eqnarray}
S & = &   \int d^4 x \left[\frac{\partial_\mu h_{ij} \partial^\mu h^{ij} }{64\pi G}\right. \nonumber \\ && + \left.\delta^3({\bf x} - {\bf M})   \frac{m_M(\delta_{ij}+h_{ij}(t,{\bf x}))}{2}  \dot M^i \dot M^j \right].
\label{eq:action}
\end{eqnarray}
The first term is identified as the usual Einstein-Hilbert action, and the second as the mirror kinetic term.  We have the freedom to fix the origin of the coordinate system at the beam splitter, with displacements relative to the beam splitter given by $M^i - B^i = X^i - \frac{h^i_k X^k}{2}$.  We also take the interferometer arm of interest aligned along the $x$ axis such that $X^x \equiv X$.  
The action Eq.~(\ref{eq:action}) then becomes
\begin{eqnarray}
S  = &  & \int dt \int \frac{d^3 k}{(2\pi)^3} \frac{m_\phi}{2} \left(\dot{\tilde \phi}(t,{\bf k})^2 - k^2 \tilde \phi(t,{\bf k})^2\right)  \\ &&+   \int dt  \frac{m_M}{2}\left(\dot{X}^2+g\int \frac{d^3 k}{(2\pi)^3} \tilde\phi(t,{\bf k})  \frac{d}{dt}(\dot{X} X)\right)  , \nonumber
\label{eq:action2}
\end{eqnarray}
where $m_\phi = 1/16 \pi G^2$, $\tilde \phi = \sqrt{G} \phi$, and $g =1/ \sqrt{G}$.   In this form, the action has been explicitly written as a non-relativistic quantum mechanical system for which the FV action is ideally suited.

Our goal is to compute the fluctuations in the mirror position $X$ driven by the gravitational $\phi-X$ coupling, with the FV influence functional. The FV influence functional is a path integral formulation for the transition probability $P_{nm}$ of some test system $Q$ from a state $\psi_n$ at time $t = \tau$ to $\psi_m $ at time $t = T$, when coupled to an interaction system characterized by an influence functional ${\cal F}(Q,Q')$:
\beq
P_{nm} & = & \int \psi_m^*(Q_T) \psi_m(Q'_T) e^{ i (S_0(Q) - S_0(Q'))} {\cal F}(Q,Q') \nonumber \\  & \times & \psi_n^*(Q_\tau) \psi_n(Q'_\tau) {\cal D} Q {\cal D} Q' dQ_T dQ'_T d Q_\tau dQ'_\tau.
\eeq
Here $S_0(Q)$ is the action in the absence of any coupling to an external system, and the influences of the external system are contained in ${\cal F}(Q,Q')$, which is characterized by the influence phase $\Phi(Q,Q')$:
\beq
{\cal F}(Q,Q') = e^{i \Phi(Q,Q')}.
\eeq

The influence phase is most simply derived for an oscillator $\tilde \phi$ coupled to the test system $Q$ through a linear interaction of the type (see Eq.~(4.6) of Ref.~\cite{Feynman:1963fq}),
\beq
S_k = \int_{\tau}^T dt \left[\frac{m_\phi}{2} (\dot {\tilde \phi}^2 - k^2\tilde \phi^2)+ \frac{g m_M}{2}\tilde\phi 
Q \right] + S_0^Q,
\label{eq:linearAction}
\eeq 
where $S_0^Q$ is the uncoupled action of the test system, and we have left a subscript $k$ explicit to denote that we have chosen a single $k$ mode.  
For this action FV showed the influence functional is
\beq
i \Phi_{k}(Q,Q') &=& - \frac{g^2 m_M^2}{8 m_{\phi} k} \int_\tau^T dt \int_\tau^t ds  (Q_t - Q'_t) \nonumber \\  & & \times \left(Q_s e^{-i k(t-s)} - Q_s' e^{i k(t-s)}\right). 
\eeq
For a distribution of oscillators $\rho(k)$ of momentum $k$ acting on the test system, we have
\beq
\Phi(Q,Q') =  \int \frac{d^3 k}{(2\pi)^3} \rho(k) \Phi_{k}(Q,Q').
\label{eq:Phitot}
\eeq

Now we decompose the influence functional into real and imaginary parts, according to
\beq
f(\Delta t)&  \equiv & \frac{g^2 m_M^2}{4 m_\phi} \int \frac{d^3 k}{(2\pi)^3} \frac{\rho(k)}{k} e^{i k \Delta t} \nonumber \\ & \equiv & A_{FV}(\Delta t) + i B_{FV}( \Delta t).
\eeq
The real and imaginary parts of the influence phase are fixed relative to each other by the oscillator density of states $\rho(k)$, which is the path integral realization of the Fluctuation-Dissipation Theorem.  The influence phase can be written in terms of this function as
\beq
i \Phi(Q,Q') &=& - \frac{1}{2} \int_\tau^T dt \int_\tau^t ds  (Q_t - Q'_t) \nonumber \\  & & \times \left(Q_s f^*(t-s) - Q_s' f(t-s) \right).
\eeq

One can see that $A_{FV}(t-s)$ represents the {\em variance} in $Q(t)$, {\em i.e.} the amplitude of the noise in the variable $Q$.  This is most clearly seen by rewriting
\beq
&& \mbox{exp}\left[\mbox{Re}\left[i \Phi(Q,Q')\right]\right] \\
&=&\mbox{exp}\left[  -  \frac{1}{2} \int_0^T dt \int_0^t ds ~A_{FV}(t-s) (Q_t - Q'_t)(Q_s - Q'_s)\right] \nonumber \\ &=&\int {\cal D }F_Q \mbox{exp}\left[- \frac{1}{2}\int_0^T dt \int_0^tds ~A_{FV}^{-1}(t-s) F_Q(t) F_Q(s)\right. \nonumber \\ &&\left.+ i \int_0^T dt  F_Q(t) (Q_t - Q'_t)\right] \nonumber,
\eeq
from which one reads
\beq
\langle F_Q^2 \rangle = A_{FV}(\Delta t).
\eeq

The only step that remains is to relate this noise to the observable, fluctuations in the position of the mirror, $X(t)$.  The influence phase has allowed us to effectively integrate out the pixellon oscillators $\phi$ and write their effect only in terms of the degree-of-freedom $F_Q$ which describes the distribution of the mirror displacements according to $Q \equiv  \frac{d}{dt}(\dot{X} X)$.   The noise $F_Q$ accordingly becomes a term in the effective action $S_{\rm eff}$ replacing the pixellon: 
\beq
S_{\rm eff} =  \int dt  \left[\frac{m_M}{2} \dot{X}^2+ F_Q(t) \frac{d}{dt}(\dot{X} X) \right].
\eeq
The equation of motion is thus
\beq
m_M \ddot X(t) =  \ddot F_Q(t) X(t),
\eeq
where we have integrated the right-hand-side by parts as necessary to move all time derivatives onto $F_Q$.   This result agrees with Ref.~\cite{Parikh:2020fhy} in the relevant limit (up to conventions), though we have arrived at it more compactly by utilizing the FV results directly.
Identifying the time-dependent change in $X(t)$ with the mirror displacement $\delta L(t)$ observed in an interferometer, we have
\beq
\frac{\delta L(t)}{L} = \frac{F_Q(t)}{m_M}, 
\eeq
or equivalently
\begin{eqnarray}
\langle \frac{\delta L^2(\Delta t)}{L^2} \rangle & = & \frac{ A_{FV}(\Delta t)}{ m_M^2} \\ & = & 4\pi G \int \frac{d^3 k}{(2\pi)^3}\frac{\rho(k)}{k} \cos (k \Delta t) dk. \nonumber 
\label{eq:LFlucsSqueezed}
\end{eqnarray}

\mysection{Observational Consequences of the Pixellon}
We now identify the pixellon with the degrees-of-freedom having density of states $\rho(k)$ in Eq.~(\ref{eq:Phitot}).  Taking $p = k$ in Eq.~\eqref{eq:rhopixapprox}, we have $\rho_{\rm pix}(k) \approx \frac{2 m_{\rm pix}}{\beta k^2}$ such that
\beq
\langle \frac{\delta L^2(\Delta t)}{L^2} \rangle \approx \frac{4 \pi G m_{\rm pix}}{\pi^2 \beta}\int_{k_{\rm min}}^{k_{\rm max}}  \frac{dk}{k}  \cos (k \Delta t),
\label{eq:delLdelt}
\eeq
where ${k_{\rm min}},~{k_{\rm max}}$ are the maximum and minimum frequencies detectable by the apparatus measuring the metric fluctuations.  
We previously argued that $m_{\rm pix}$ is near the Planck scale; ultimately it should be experimentally determined. Taking, for example, $2 m_{\rm pix}/\beta = a/l_p L$, and in the limit $\Delta t \rightarrow 0$, we recover
\beq
\langle \frac{\delta L^2}{L^2} \rangle \approx \frac{a l_p}{4\pi^2 L} \log\left(\frac{ k_{\rm max}}{k_{\rm min}}\right).
\label{eq:delLpix}
\eeq
Now $k$ is the momentum conjugate to the spacetime position $x$.  Thus we can identify $~k_{\rm min}$ with the overall apparatus size $k_{\rm min} \sim 1/L$, and the spatial separation between the two arms of the interferometer $k_{\rm max} \sim 1/| {\bf r}_1- {\bf r}_2|$.  Note that this is remarkably similar to the result obtained in Ref.~\cite{Verlinde:2019xfb} despite the distinct {\em Ansatz} and calculational techniques.  

The amplitude of the effect is within reach.  For example, the interferometer of Ref.~\cite{Chou:2017zpk} was able to constrain a power spectral density $S_h(\omega) \equiv \int_{-\infty}^\infty d\Delta t \langle \frac{\delta L^2(\Delta t)}{L^2} \rangle e^{-i\omega \Delta t} \lesssim 0.15~l_p/\sqrt{8 \pi} (\sinc^2(\omega L/2) - 2 \sinc^2(\omega L))$.  While a precise phenomenological analysis comparing our model to the data is not the purpose of this present work, we can obtain an estimate by integrating Eq.~(\ref{eq:delLdelt}) times $e^{-i\omega \Delta t} $ over $-L$ to $L$, taking $\omega \rightarrow 0$.  With this naive estimate, we find that the data is roughly consistent with $a \lesssim 2$.  This motivates future work to make precise phenomenological predictions and to search for the unique signatures of this model.  For example, Ref.~\cite{Verlinde:2019xfb} went one step further and used the spherical symmetry of the entangling surface to argue that length fluctuations correlated between two interferometer arms should obey the Green function of the 2-d Laplacian on the sphere.  While we have not repeated this calculation here, the general form of Eq.~(\ref{eq:delLpix}) suggests similar behavior holds for this model.  The observation of these angular correlations would be a spectacular confirmation of the idea of low momentum vacuum fluctuations in quantum gravity, and could be tested with a suitable modification of a planned table top interferometer experiment \cite{Vermeulen:2020djm}.  We anticipate a vibrant experimental program along these directions. 

{\em Acknowledgments.}
I especially thank Tom Banks, Cindy Keeler, Michele Papucci and Julio Parra-Martinez for detailed comments on the draft.  I thank many others for discussion and criticism, including Dan Carney, Yanbei Chen, Clifford Cheung, Patrick Draper, David B. Kaplan, Dongjun Li, Juan Maldacena, Maulik Parikh, and Erik Verlinde.

\bibliography{QG}

\end{document}